\documentclass[sigconf]{acmart}

\AtBeginDocument{%
  \providecommand\BibTeX{{%
    \normalfont B\kern-0.5em{\scshape i\kern-0.25em b}\kern-0.8em\TeX}}}

\setcopyright{acmcopyright}
\copyrightyear{2022}
\acmYear{2022}
\acmDOI{.}

\acmConference[ACM]{ACM}{2022}{Virtual Event}

\usepackage{amsmath}  
  {
      \theoremstyle{plain}
      \newtheorem{assumption}{Assumption}
  }
\usepackage{color}
\usepackage{algorithmic}
\usepackage{graphicx}
\usepackage{textcomp}
\usepackage{xcolor}
\usepackage{multirow}
\usepackage{amsthm}
\usepackage{subfig}
\usepackage{array}  
\begin{document}

{\color{red}\title{A Fully Decentralized Tuning-free Inexact Projection Method for P2P Energy Trading}}
\author{Meiyi Li}
\email{meiyil@utexas.edu}
\orcid{0000-0002-0178-7883}
\affiliation{%
  \institution{The University of Texas at Austin}
  \city{Austin}
  \state{Texas}
  \country{USA}
}

\author{Javad Mohammadi}
\email{javadm@utexas.edu}
\affiliation{%
  \institution{The University of Texas at Austin}
  \city{Austin}
  \state{Texas}
  \country{USA}
}

\author{Soummya Kar}
\email{soummyak@andrew.cmu.edu}
\affiliation{%
  \institution{Carnegie Mellon University}
  \city{Pittsburgh}
  \state{Pennsylvania}
  \country{USA}
}

\renewcommand{\shortauthors}{author.}

\begin{abstract}
Agent-based solutions lend themselves well to address privacy concerns and the computational scalability needs of future distributed electric grids and end-use energy exchanges. Decentralized decision-making methods are the key to enabling peer-to-peer energy trading between electricity prosumers. However, the performance of existing decentralized decision-making algorithms highly depends on the algorithmic design and hyperparameter tunings, limiting applicability. This paper aims to address this gap by proposing a decentralized inexact projection method that does not rely on parameter tuning or central coordination to achieve the optimal solution for Peer-to-Peer (P2P) energy trading problems. The proposed algorithm does not require parameter readjustments, and once tuned, it converges for a wide range of P2P setups. Moreover, each prosumer only needs to share limited information  (i.e., updated coupled variable) with neighboring prosumers. The IEEE 13 bus test system is used to showcase our proposed method's robustness and privacy advantages.
\end{abstract}

\ccsdesc[500]{Hardware~Smart grid}
\ccsdesc[500]{Computing methodologies~Distributed algorithms}

\keywords{P2P energy trading, distributed algorithm}


\maketitle
\section*{Nomenclature}

\begin{table}[htbp]
\begin{tabular}{lp{6cm}}
$\mathcal{N_{P}},i,j$                                             & Set and index of ${N_{P}}$ prosumers                                                                       \\
$\mathcal{N}_{PL}^{i},\mathcal{N}_{ESS}^{i},\mathcal{N}_{DE}^{i}$ & Set of Prosumer $i$'s flexible loads, energy storage systems and diesel engines                               \\
$m,q,p$                                                           & Indexes associated with flexible loads, energy storage systems, and diesel engines                                            \end{tabular}
\end{table}
    
\begin{table}[htbp]
\begin{tabular}{lp{6.3cm}}

$P_{IL,t}^{i},P_{FL}^{i}( t)$                                     & Total cosuming power of Prosumer $i$'s inflexible and flexible loads at time $t$  \\                             $P_{NG,t}^{i},P_{DG}^{i}( t)$                                     & Total generation power of Prosumer $i$'s non-dispatchable \& dispatchable generation at $t$  \\
$P_{o}^{i} (t)$                                                   & Prosumer $i$'s output at time $t$                                                 \\
$P_{FL}^{i,m}( t)$                                     & Total cosuming power of Prosumer $i$'s $m$th flexible loads at time $t$   \\
$W_{FLref}^{i,m}$                                                 &Minimal consuming energy requirement of Prosumer $i$'s $m$th flexible load for the day                              \\
$\beta_{PL2}^{i,m},\beta_{PL1}^{i,m}$                             &  Sensitivity to load shifting and the total power consumption of Prosumer $i$'s $m$th flexible load\\
$S^{i}$                                                           & Prosumer $i$'s convenience function for flexible loads                                                    \\
$P_{C}^{i,q}(t),P_{D}^{i,q}(t)$                                   & Charging and discharging power of Prosumer $i$'s $q$th energy storage system at time $t$                     \\
$P_{Cmax}^{i,q},P_{Dmax}^{i,q}$                                   & Upper bound of charging and discharging power                 \\
$\lambda_{ESS}^{i,q}$&Prosumer $i$'s operation fees coefficient for $q$th energy storage system \\
$\eta _{C}^{i,q},\eta _{D}^{i,q}$                                 & Charging and discharging efficiencies   \\
$W_{ESS0}^{i,q},W_{ESSN}^{i,q}$                                   & Initial capacity and nominal capacity of energy storage system   \\
$P_{DE}^{i,p}(t)$             & Generation power of Prosumer $i$'s $p$th diesel engine            \\
$P_{DEmax,t}^{i,p}$ & Upper bound of generation power of diesel engine                         \\
$P_{DEmin,t}^{i,p}$ & Lower bound of generation power of diesel engine                        \\
$R_{max,t}^{i,p},R_{min,t}^{i,p}$                                 & Upper and lower bound for ramping power of diesel engine                              \\
$\lambda_{DE1}^{i,p},\lambda_{DE2}^{i,p}$&Operation fees coefficient for diesel engine $p$ of prosumer $i$\\
$F^i_{om}$                                                        & Prosumer $i$'s operating and maintenance cost of energy storage systems and diesel engines                \\

$F_{grid}^{i}$, $F_{trade}^{i}$                                   & Prosumer $i$'s cost related to the trades with the grid and other prosumers                                \\
$F_{o}^{i}$                                                       & Prosumer $i$'s operation fees for the system and the electrical distance cost  \\
$P_{gs}^{i}( t),P_{gb}^{i}( t)$ & Prosumer $i$'s selling and buying power from and to grid\\
$P_{ps}^{i,j}(t),P_{pb}^{i,j}(t)$ & Prosumer $i$'s selling and buying power from and to Prosumer $j$\\
$\lambda _{gs},\lambda _{gb}$&Selling and buying price with the grid\\
$\lambda_{p}^{i,j}(t)$ &Trading price between Prosumer $i$ and $j$\\
$\lambda _{max},\lambda _{min}$&Upper and lower bound of trading price among prosumers 
\\

$\lambda _{o}$ & System operation fees coefficient \\
$\lambda _{d}d^{i,j}$ & Electrical distance cost between Prosumers $i$ and $j$\\
$\mathcal{L}$, $l$ &Set and index of lines in the community\\
$Y_{l}^{i}$&Relationship coefficient of Prosumer $i$ and Line $l$\\
$C_{min}^{l},C_{max}^{l}$&Upper and lower limits for line flow of Line $l$\\
\end{tabular}
\end{table}

\begin{table}[htbp]
\begin{tabular}{lp{6.5cm}}

$\mathbf{X}$&Vector of all the stacked variables of power\\
$\boldsymbol{\lambda}$ &Vector of all the stacked variables of trading prices\\
$\mathbf{X}^{i}_{c}$, $\mathbf{X}^{i}_{u}$& Prosumer $i$'s coupled \& uncoupled variable vectors\\
$\mathbf{X}^{*},\mathbf{X}_{0}^{*}$&Optimal solution with and without energy trading\\
$k,d$& Index for gradient descent and inner iteration\\
${x}_{r},\mathcal{N}_{c}^{r}$&$r$th element in $\mathbf{X}$ and its set of neighbors
\end{tabular}
\end{table}

\section{Introduction}

The future of electric power grids is distributed \cite{7581042}; hence, management responsibilities will be shared between multiple entities (agents) \cite{8973917}. Although these agents are physically interconnected, they may pursue different goals. Depending on the electric demand, availability of self-generation, and electricity prices, each agent may collaborate or compete to achieve the best individual outcome. P2P energy trading \cite{7422059} has become a driving force for enabling intra-agent energy exchanges and is paving the way for the transition to a multi-agent electric grid. Peer-to-Peer energy trading allows end-users to share their excess energy, making the economics of Distributed Energy Resource (DER)s more attractive. Energy trading often takes place across a local distribution system and, if appropriately managed, can help alleviate congestion management.


The P2P energy trading models can be cast as optimization problems where variables of each prosumer are tied through coupling constraints. 
Existing studies on multi-agent scheduling and bidding in P2P energy trading setups can be clustered into three optimization classes; cooperative\cite{8957839, LONG2018261}, competitive\cite{8928505, 9049140,8490840}, and hybrid strategies\cite{8316917,7422059}. 
These multi-agent frameworks are also well-suited to address the growing privacy concerns as subproblems of the original problem are often solved by individual agents \cite{9220870}. The Alternating Direction Method of Multipliers (ADMM) is the most common method for solving energy trading problems in a distributed fashion\cite{8356100, 8669964, 8409328}. Authors in \cite{8356100} have used ADMM to devise a distributed price-directed optimization mechanism for improving scalability and preserving prosumers' privacy. The closed-form solutions to all sub-problems are derived in \cite{8409328}  to improve the computational efﬁciency of ADMM. Moreover, \cite{8669964} uses a fast ADMM approach to minimize the energy cost of buildings' operation. The underlying distributed mechanism of these studies requires a central coordinator to update and disseminate the Lagrangian multipliers. Each control entity only communicates with the coordinator,
and there are no direct communication links between the entities.

On the other hand, the decision-making hierarchy of fully decentralized methods is flat and does not rely on a central entity. In this regard,  \cite{8630697} used a consensus-based ADMM method to enable energy trading negotiations between autonomous prosumers capable of P2P information exchange.
The primal-dual gradient methods and consensus-based approaches are also commonly used for decomposing energy trading problems into regional sub-problems. For example, authors in \cite{8957839} and \cite{8782819} proposed trading schemes for P2P trading using KKT optimality conditions to update dual variables. Also, \cite{8481579} presented a relaxed consensus + innovation approach to solve the energy trading problem in a fully decentralized manner.


The performance of the discussed methods relies on hyperparameters such as Lagrangian multipliers.  
Due to scalability needs and privacy concerns, tuning these parameters while preserving privacy in a practical P2P energy market setup is burdensome. Put differently, these methods require different tuning parameters for dissimilar optimization setups. In some cases, new tuning parameters should be adopted even with a minor problem reformulation. Therefore, the robustness of decentralized optimization algorithms with respect to tuning parameters is critical for practical power grid optimization problems.

This paper proposes a fully decentralized and parameter tuning-free scheduling method to solve P2P energy trading problems. We use a two-stage energy trading strategy in \cite{8669964} as the P2P setup where prosumers first cooperate to determine the quantity of traded energy and internally compete to determine the trading price afterward. The main contributions include:

\begin{itemize}
\item  

Our method adopts a fully decentralized projected gradient descent algorithm. Therefore, it does not need a central coordinator for regulating the information processing procedure. The proposed method only requires each prosumer to share updated coupled variables with corresponding neighbors. Also, our solution needs limited information sharing, hence, preserving prosumers' privacy.

\item 

The convergence of the proposed method does not rely on tuning parameters, and the algorithm works for a wide range of similar problems without changing parameters. This convergence property is analytically justified in this paper. In addition, we analyze the effects of parameter design on the performance (i.e., convergence rate) of the proposed method. 

\item 

The proposed method does not need slack variables or the Lagrangian multiplier to accommodate for inequalities constraints of original optimization problems in the decentralized decision-making procedure. This reduces the computational burden of agent-based computations and improves the scalability of the decentralized decision-making method.

\end{itemize}


Note, while the proposed method is applied to solve P2P energy trading problems, it can solve a broad range of problems, including; energy management of smart buildings, demand response in microgrids, coordination control of inverter-based distributed generation.


The paper is organized as follows: Section II presents the system model. Section III presents the P2P energy trading setup. The proposed decentralized inexact projection solution method is discussed in Section IV. Finally, the contributions of this work are showcased using the IEEE 13-bus case study.

\section{System model}
\begin{figure}[htbp]

\centering
\includegraphics[width=0.99\columnwidth]{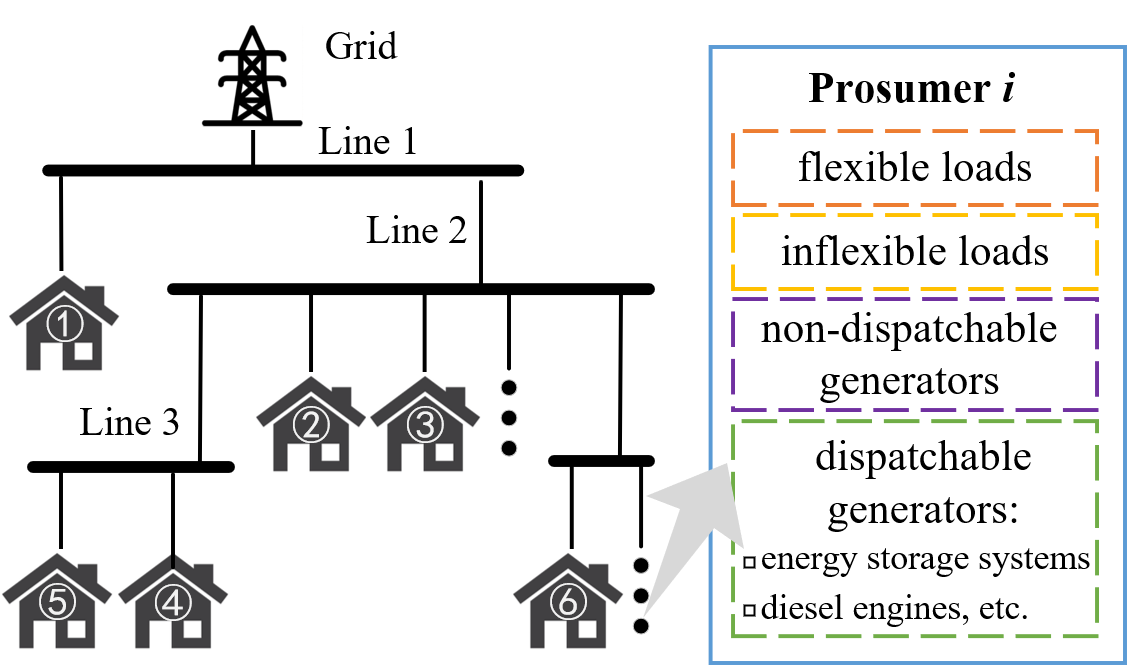}
\caption{P2P energy trading community.}
\label{f:community}
\end{figure}

In this paper, we consider a P2P energy sharing community as shown in Figure \ref{f:community}. The time interval of the scheduling process is considered as $T_{loop}=24h$. Line loss is ignored in our analysis. In what follows, we discuss modeling specifics for different system components.

\subsection{Generators and loads}
We consider four distinct models for generators and loads of prosumers; Non-dispatchable Generators (NGs), Inflexible Loads (ILs), Flexible Loads (FLs), and dispatchable generators, e.g., Energy Storage Systems (ESSs) and Diesel Engines (DEs).  
We assume that the power of inflexible loads  and non-dispatchable generation  are accurately predicted. 
As for flexible loads  and dispatchable generators, the modeling details are presented below. 
\begin{align}
&P_{FLmin,t}^{i,m}\leq P_{FL}^{i,m}(t) \leq P_{FLmax,t}^{i,m}\label{CPPL} \\
&\sum_t P_{FL}^{i,m}( t)\Delta t\geq W_{FLref}^{i,m}\label{CPPLW}\\
 SOC_{min}^{i,q}\leq& \frac{W_{ESS0}^{i,q}+\sum_{\tau=\Delta t}^{t} ( P_{C}^{i,q} ( \tau  )\eta _{C}^{i,q}-\frac{P_{D}^{i,q}(\tau )}{\eta _{D}^{i,q}} )\Delta t}{W_{ESSN}^{i,q}}\leq SOC_{max}^{i,q}
\label{ESSSOC}\\
 0\leq& P_{C}^{i,q}(t)\leq P_{Cmax}^{i,q},\textup{ }0\leq P_{D}^{i,q}(t)\leq P_{Dmax}^{i,q}\label{CPC}\\
W_{min}^{i,q}\leq& 
\sum_{t}\left ( P_{C}^{i,q} (t)\eta _{C}^{i,q}-\frac{P_{D}^{i,q}(t)}{\eta _{D}^{i,q}} \right )\Delta t\leq W_{max}^{i,q}\label{ESSEND} \\
 &P_{DEmin,t}^{i,p}\leq P_{DE}^{i,p}(t)\leq P_{DEmax,t}^{i,p}\label{CPDE}
\\
\Delta t R_{min,t}^{i,p}& \leq P_{DE}^{i,p}(t)- P_{DE}^{i,p}\left ( t-\Delta t \right )\leq \Delta t R_{max,t}^{i,p} \label{CRDE}   
\end{align}

 \noindent Hence, the net output power of prosumer i can be presented as: 
\begin{align}
P_{o}^{i} (t)=&P_{NG,t}^{i}+\sum_{p\in \mathcal{N}_{DE}^{i}}  P_{DE}^{i, p}(t)+\sum_{q\in \mathcal{N}_{ESS}^{i}} (P_{C}^{i,q}(t)-P_{D}^{i,q}(t))\nonumber\\
&-P_{IL,t}^{i}-\sum_{m\in \mathcal{N}_{PL}^{i}}P_{FL}^{i,m}(t)    
\end{align}

The utility function of generators and loads include 1) $F_{om}^{i}$, i.e., operating and maintenance cost of ESSs and diesel engines; 2) $S^{i}$, i.e., flexible load's convenience function. To simplify the analysis, we ignore the constant part of the quadratic cost functions.
\begin{align}
F_{om}^{i}&=\sum_{t}(\sum_{q\in \mathcal{N}_{ESS}^{i}} \lambda _{ESS}^{i,q} (P_{D}^{i,q}(t)+P_{C}^{i,q}(t))   \nonumber\\
&+\sum_{p\in \mathcal{N}_{DE}^{i}} ( \lambda_{DE1}^{i,p}P_{DE}^{i, p}(t)^{2}+\lambda_{DE2}^{i,p}P_{DE}^{i,p}(t)))
\end{align}
\begin{align}
 S^{i}&=\sum_{m\in \mathcal{N}_{PL}^{i}}\beta_{PL1}^{i,m}(W_{FLref}^{i,m}-\sum_t P_{FL}^{i,m}\left ( t  \right )\Delta t)  \nonumber \\
&-\sum_{m\in \mathcal{N}_{PL}^{i}}\sum_{t=\Delta t }\beta_{PL2}^{i,m}(P_{FL}^{i,m}(t)-P_{FLref,t}^{i,m})^2
\end{align}

\subsection{Exogenous cost}
\subsubsection{Trade costs:}
The cost related to the power exchange with the grid ($F_{grid}^{i}$) and other prosumers ($F_{trade}^{i}$) are:
\begin{align}
 &F_{grid}^{i}=\sum_t (\lambda _{gb}P_{gb}^{i}(t)-\lambda _{gs}P_{gs}^{i}(t)) \\
 F_{trade}^{i}&=\sum_t \sum_{j\in \mathcal{N_{P}}/i}\lambda _{p}^{i,j}( t)(P_{pb}^{i,j}(t)-P_{ps}^{i,j}(t))
\end{align}

\noindent These costs are subject to the following constraints:

\begin{align}
    &\lambda _{p}^{i,j}(t)=\lambda _{p}^{j,i}(t)\label{Cprice}\\
    &\lambda _{min}\leq \lambda _{p}^{i,j}( t)\leq\lambda _{max}\label{Lprice}\\
   &P_{gs}^{i}( t)\geq0 , P_{gb}^{i}( t)\geq0 \label{CPg} 
\\
&P_{ps}^{i,j}( t)\geq 0, P_{pb}^{i,j}( t)\geq 0\label{CPp}\\
   &P_{pb}^{i,j}(t)=P_{ps}^{j,i}(t)\label{CPpsb}   \\
   \label{Po}
P_{o}^{i} (t)=P_{gs}^{i}(t)&-P_{gb}^{i}(t)+\sum_{j\in \mathcal{N_{P}}/i}(P_{ps}^{i,j}(t)-P_{pb}^{i,j}(t))
\end{align}

Note (8) and (18) preserve the equality between net output of Prosumer $i$ and it's traded power.

\subsubsection{System's operation fees and the electrical distance cost}
The system collects the operation fees to cover the operation expenses, and the electrical distance cost incentivizes prosumers in the community to trade with their electrically-closest prosumers \cite{8630697}.  These cost are captured as: 
\begin{align}
\label{eq:electrical_d}
  F_{o}^{i}(t)= ( \lambda _{o}+\lambda _{d}d^{i,j})\sum_{j\in \mathcal{N_{P}}/i}(P_{pb}^{i,j}(t)+P_{ps}^{i,j}(t))  
\end{align}

\subsection{Network congestion constraints}
 We use $Y_{l}^{i}$ to show the direct relationship between Prosumer $i$ and the line $l, l\in \mathcal{L}$. Here, $Y_{l}^{i}=0$ if the active power of the line $l\in \mathcal{L}$ is not determined by $P_{o}^{i}(t)$ and  $Y_{l}^{i}=1$ for the otherwise, e.g., $Y_{l=2}^{i=2}=1$ and $Y_{l=2}^{i=1}=0$ in Figure \ref{f:community}. Then, for those lines $l$ $s.t.Y_{l}^{i}=1$, we have: 

\begin{equation}
 C_{min}^{l}\leq P_{o}^{i}(t)+ \sum_{j\in \mathcal{N_{P}}/i}Y_{l}^{j}P_{o}^{j}(t)\leq C_{max}^{l}\label{Cline} 
\end{equation}

\section{Problem formulation}

\begin{figure}[htbp]
\centering
\includegraphics[width=0.95\columnwidth]{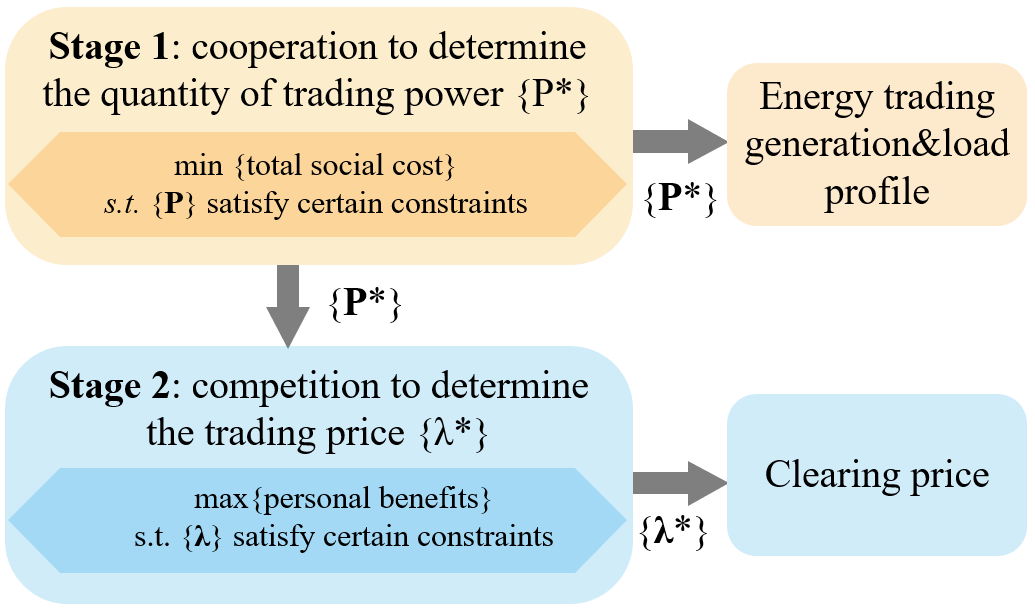}
\caption{two-stage energy sharing strategy as proposed in \cite{8669964}.}
\label{f:strategy} 
\end{figure}

Authors in \cite{8669964} proposed a two-stage energy sharing strategy to faciliate energy sharing among smart buildings. As shown in Figure \ref{f:strategy}, the optimal energy trading profile is determined by minimizing the total social cost in the first stage. Then, the optimal energy trading profile is used as the input to determine the trading price (through a competitive process) in the second stage. In this paper, we use the discussed two-stage energy sharing strategy, which is also shown in Figure \ref{f:strategy}.
We will first formulate the problem based on the system model in section II. Later in section 4, we will use a fully decentralized inexact projection method to solve the problem. 

\subsection{Optimal energy trading profile}
The total trading cost of a prosumer is given as:
\begin{align}
  f^{i}=F_{om}^{i}-S^i+F_{grid}^{i}+F_{o}^{i}+F_{trade}^{i}  
\end{align}

Hence, the cost of all prosumers adds up to
\small $f= \sum_{i\in \mathcal{N_{P}}}f^{i}$ \normalsize. 
The set of constraints for all prosumers is a collection of individual constraints, i.e.,

 \begin{align}
\mathbb{S}^{i}=\left \{ \eqref{CPPL} \right.-\eqref{CRDE},\eqref{CPg}-\eqref{Po}, \eqref{Cline} \left.\right \} , \mathbb{S}=\bigcap_{i\in \mathcal{N_{P}}}\mathbb{S}^{i} \label{define_si} 
\end{align}

Let $\mathbf{X}$ denote the vector of all the stacked variables \small [$(P_{PL}^{i,m}(t)$, $P_{D}^{i,q}(t)$, $ P_{C}^{i,q}(t)$, $P_{DE}^{i,p}$, $P_{gs}^{i}(t)$, $P_{gb}^{i}(t))$, $P_{ps}^{i,j}(t)$, $P_{pb}^{i,j}(t)$, $P_{o}^{i}(t)$]\normalsize and $\mathbf{X}^i$ be the vector of all the stacked variables of Prosumer $i$. Then, the problem of optimal energy trading profile summarizes to:

\begin{align}
\underset{\mathbf{X}\in \mathbb{S}}{\min f(\mathbf{X})}\label{problem1} 
 \end{align}

Moving forward, we refer to $\mathbf{X}^{*}$ as the optimal solution to this problem. Then, $\mathbf{X}^{*}$ will be used as the input in the second stage to determine the trading price by competition.
\subsection{Clearing price}

Each rational prosumer aims to minimize energy purchasing costs or maximize profits. Therefore, prosumers are in competition to determine the final trading prices.  Because the energy trading relationship is determined by solving problem (\ref{problem1}), sellers would like to trade at a higher price (no higher than $\lambda _{max}$) and buyers at a lower price (no lower than $\lambda _{min}$). That is, buyers want to obtain a price as close as possible to the lowest price, and the opposite for sellers. The objective function of prosumer i would be as below. Here, $D$ is the distance.
\begin{align}
    f_{\lambda}^i=\sum _t\sum _{j\in \mathcal{N_{P}}/i}\left ( \underset{\textup{if } P_{ps}^{i,j}(t)>0}{\underbrace{D\left ( \lambda _{p}^{i,j}(t),\lambda _{max}\right )}}+\underset{\textup{if } P_{pb}^{i,j}(t)>0}{\underbrace{D\left ( \lambda _{p}^{i,j}(t),\lambda _{min}\right )}} \right )
\end{align}


Then, the objective function of all prosumers adds up to \small$f_{\lambda}= \sum_{i\in \mathcal{N_{P}}}f^{i}$ \normalsize. The energy trading should result in economic gains for prosumers. That is, each prosumer should spend less compared to the without energy trading case: 
\begin{align}
    f^i(\mathbf{X}^{i*})\leq f^i(\mathbf{X}_0^{i*})\label{pricebasic}
\end{align}
Here $\mathbf{X}_0^{i*}$ is the optimal power profile of problem (\ref{problem1}) by adding constraints \small
$P_{ps}^{i,j}(t)=P_{pb}^{i,j}(t)=0, \forall i,j\in \mathcal{N_{P}}$.\normalsize That is, $\mathbf{X}_0^{i*}$ is the optimal solution of power without energy trading.

The set of constraints for prices is:

\begin{align}
\mathbb{S}_{\lambda}^{i}=\left \{ \eqref{Cprice},  \eqref{Lprice},\eqref{pricebasic} \right \} , \mathbb{S}_{\lambda}=\bigcap_{i\in \mathcal{N_{P}}}\mathbb{S}_{\lambda}^{i} \label{define_s_lambdai} 
\end{align}

Let $\boldsymbol{\lambda}$ denote the vector of all the stacked variables  [$\lambda_{p}^{i,j}(t)$]. Then, the
problem of optimal price summarizes to:

\begin{align}
\underset{\boldsymbol{\lambda}\in \mathbb{S_{\lambda}}}{\min f_{\lambda}(\boldsymbol{\lambda})}\label{problem2} 
 \end{align}

\section{Fully decentralized inexact projection method}

\begin{figure}[htbp]
\centering
\includegraphics[width=0.95\columnwidth]{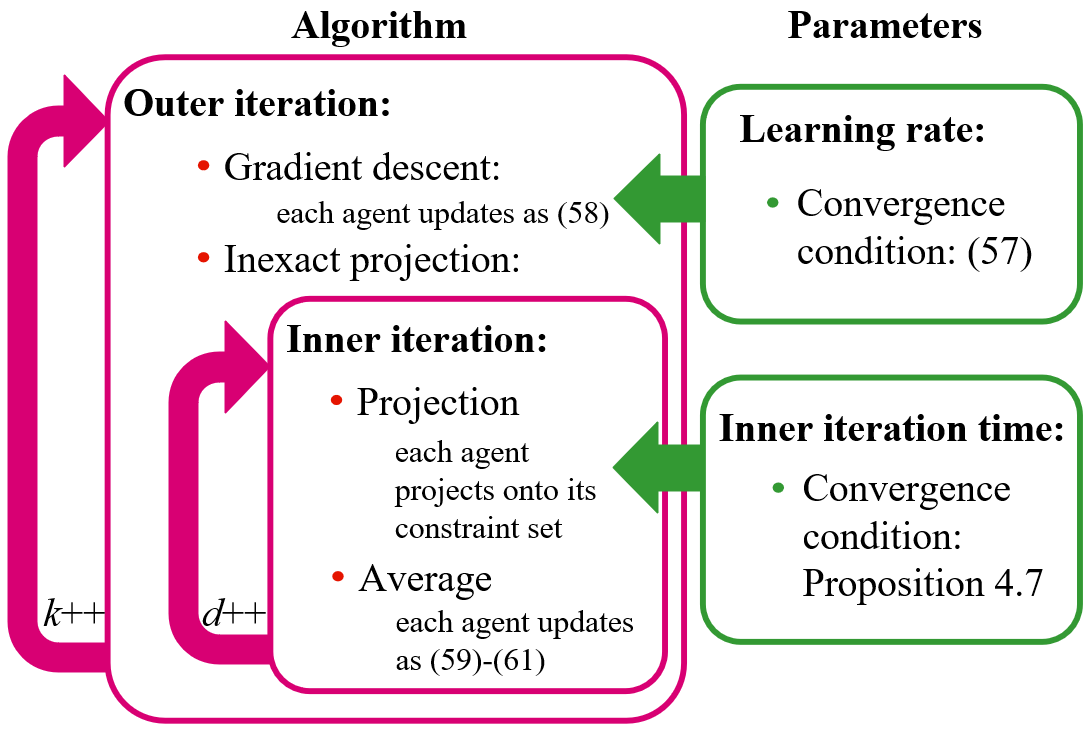}
\caption{Decentralized inexact projection method.}
\label{f:method}
\end{figure}

 As shown in Figure \ref{f:method}, we adopt an inner-outer iteration method based on inexact projected gradient descent to solve the earlier described problem (\ref{problem1}) and (\ref{problem2}). Since problem (\ref{problem1}) and (\ref{problem2}) share the same form, we will only use the expression in problem (\ref{problem1}) to discuss the method. The algorithm and its convergence analysis are also applicable to problem (\ref{problem2}).
\subsection{Inner-outer algorithm}
The outer iteration (with index $k$) executes gradient descent, whereas the inner iteration (with index $d$) performs projection. 
 The outer iteration uses the inexact projected gradient descent method as presented in (\ref{GD}) and (\ref{projection}):

\begin{align}
    &\mathbf{\widetilde{X}}^{[k]}=\mathbf{X}^{[k]}-\frac{1}{L}\triangledown f(\mathbf{X}^{[k]})\label{GD}\\
    \textup{compute }&\mathbf{X}^{[k+1]}\textup{, s.t.}\left \| \mathbf{X}^{[k+1]}-\mathbb{P_{S}}(\mathbf{\widetilde{X}}^{[k]}) \right \|\leq \varepsilon _{proj}^{[k]}\label{projection}
\end{align} 

Here $1/L$ is the learning rate. To ensure that  $\mathbf{X}^{[k+1]}$ satisfies (\ref{projection}), we use the averaged projection method in \cite{5404774} as the inner algorithm to compute $\mathbf{X}^{[k+1]}$. Prosumer $i$ only needs to project onto his constraint set $\mathbb{S}^{i}$. This process will continue when all the prosumers reach a consensus on the power profile, i.e.,

\begin{align}
   \mathbf{X}^{[d]}_{\mathbb{S}^{i}}= \mathbb{P_{S}}_{^{i}}(\mathbf{w}^{[d]} ),\forall i\in \mathcal{N_{P}},d\geq 0\label{distributed projection}\\
   \mathbf{w}^{[d+1]}=\frac{1}{N_{P}}\sum_{i\in \mathcal{N_{P}}}\mathbf{X}^{[d]}_{\mathbb{S}^{i}},d\geq 0
   \label{average projection}
\end{align}

Here $\mathbf{w}^{[d=0]}=\mathbf{\widetilde{X}}^{[k]}$. Let inner iteration  (\ref{distributed projection}) and (\ref{average projection}) perform for $N_{inner}^{[k]}\geq 1$ iteration which results in $\mathbf{X}^{[k+1]}=\mathbf{w}^{[N_{inner}^{[k]}]}$.
\subsection{Convergence analysis}
We present several proofs to show the convergence of the proposed method. First, we have Lemma (\ref{lemma:inner}) according to \cite{5404774}.
\begin{lemma}
\label{lemma:inner}
 Given nonempty closed convex sets $\mathbb{S}^{i}\in \mathbb{R}^{n}, i = 1,...,N_{P}$, let $\mathbb{S}=\bigcap_{i\in \mathcal{N_{P}}}\mathbb{S}^{i}$ denote their intersection,  $\mathbb{S}$ is nonempty.  Let $\mathbf{X}^{[d]}_{\mathbb{S}^{i}}$ and $\mathbf{w}^{[d]}$ be deﬁned by (\ref{distributed projection}) and (\ref{average projection}). Then, we have the following. 

(a) \begin{align}
 \underset{d\rightarrow \infty }{\textup{lim}}\left \| \mathbf{w}^{[d]}-\mathbb{P_{S}}(\mathbf{w}^{[d=0]}) \right \|_{2}=0  
\end{align}

(b) For all $d\geq 0$, and $\forall \mathbf{Z}\in \mathbb{S}$, we have:

\begin{align}
\left \| \mathbf{w}^{[d+1]} -\mathbf{Z} \right \|_{2}^{2}\leq   \left \| \mathbf{w}^{[d]}-\mathbf{Z}\right \|_{2}^{2}-\frac{\sum_{i\in \mathcal{N_{P}}}\left \| \mathbf{w}^{[d]}-\mathbb{P_{S}}_{^{i}}(\mathbf{w}^{[d]})  \right \|_{2}^{2}}{N_{P}}  
\end{align}

(c) 
\begin{align}
   \left \| \mathbf{w}^{[d]}-\mathbb{P_{S}}(\mathbf{w}^{[d=0]}) \right \|_{2}\leq 2\left \| \mathbf{w}^{[d]}-\mathbb{P_{S}}(\mathbf{w}^{[d]}) \right \|_{2} \label{wdwd0}
\end{align}
\end{lemma}
    
Then, we introduce the concept of "linearly regular" according to \cite{bauschke1996projection}.

\begin{lemma}
\label{lemma:lg}
 We say that the $N_{P}-$tuple of closed convex $(\mathbb{S}^{1}...\mathbb{S}^{N_{P}})$, $\mathbb{S}^{i}\in \mathbb{R}^{n}, i = 1,...,N_{P}$ is linearly regular if~$\exists \alpha>0$, $\forall  \mathbf{X}$, $\left \| \mathbf{X}-\mathbb{P_{S}}(\mathbf{X})  \right \|_{2}\leq \alpha  \max\left \{ \left \| \mathbf{X}-\mathbb{P_{S}}_{^{i}}(\mathbf{X}) \right \|_{2}, i\in\mathcal{N_{P}}  \right \}$.
 Here, $\mathbb{S}=\bigcap_{i\in \mathcal{N_{P}}}\mathbb{S}^{i}$ denotes their intersection,  $\mathbb{S}$ is nonempty. If each set $\mathbb{S}^{i}$ is a polyhedron, then the tuple $(\mathbb{S}^{1}...\mathbb{S}^{N_{P}})$ is linearly regular.
\end{lemma}
 
Then, we start the analysis by giving two assumptions.

\begin{assumption}
\label{assumption:si_nemp}
Given  sets $\mathbb{S}^{i}\in \mathbb{R}^{n}, i = 1,...,N_{P}$ defined by (\ref{define_si}), and $\mathbb{S}=\bigcap_{i\in \mathcal{N_{P}}}\mathbb{S}^{i}$ denotes their intersection,  $\mathbb{S}$ is nonempty.
\end{assumption}

\begin{assumption}
\label{assumption:f_smooth}
$f(\mathbf{X})$ is a $L-$smooth function.
\end{assumption}
Therefore, we have proposition \ref{proposition:inner} for the convergence rate of the inner iteration as below.
\begin{proposition}
\label{proposition:inner}
 Let Assumption \ref{assumption:si_nemp} hold. Given a sequence $\mathbf{w}^{[d]}$ generated by (\ref{distributed projection}) and (\ref{average projection}). Then, $\forall \varepsilon>0$, after at most $\overline{N}$ steps, we could obtain a  $\mathbf{w}^{[d]}$ satisfying: $$ \left \| \mathbf{w}^{[d]}-\mathbb{P_{S}}(\mathbf{w}^{[d=0]}) \right \|_{2}\leq\varepsilon$$.

where $$\overline{N}=2\frac{\log \varepsilon-\log \left \| \mathbf{w}^{[d]}-\mathbb{P_{S}}(\mathbf{w}^{[d=0]}) \right \|_{2}}{\log \eta},\frac{1}{2}\le \eta< 1$$

\end{proposition}

\begin{proof} Because each set $\mathbb{S}^{i}$ is a polyhedron according to (\ref{define_si}),
according to Lemma \ref{lemma:lg},  $\exists \alpha>0$ satisfying: 
\begin{align}
    \left \| \mathbf{w}^{[d]}-\mathbb{P_{S}}(\mathbf{w}^{[d]}) \right \|_{2} \leq \alpha \left \| \mathbf{w}^{[d]}-\mathbb{P_{S}}_{^{i}}(\mathbf{w}^{[d]}) \right \|_{2},\forall i\in\mathcal{N_{P}}
\end{align}
Here, $\mathbb{S}\subseteq\mathbb{S}^{i}$, 
we have:
\begin{align}
   \left \| \mathbf{w}^{[d]}-\mathbb{P_{S}}_{^{i}}(\mathbf{w}^{[d]})\right \|_{2} \leq  \left \| \mathbf{w}^{[d]}- \mathbb{P_{S}}(\mathbf{w}^{[d]}) \right \|_{2},\forall i\in\mathcal{N_{P}} 
\end{align}
Therefore, $\alpha\geq 1$ and $\frac{1}{2}\le \left ( 1-\frac{1}{2\alpha } \right )< 1$.

Combine with Lemma \ref{lemma:inner}(c):
\begin{align}
    \frac{1}{2 \alpha}\left \| \mathbf{w}^{[d]}-\mathbb{P_{S}}(\mathbf{w}^{[d=0]}) \right \|_{2}\leq  \frac{\sum_{i\in\mathcal{N_{P}}}\left \| \mathbf{w}^{[d]}-\mathbb{P_{S}}_{^{i}}(\mathbf{w}^{[d]}) \right \|_{2}}{N_{P}}
\end{align}

Also, according to Lemma \ref{lemma:inner}(b):
\begin{align}
&\left \| \mathbf{w}^{[d+1]} -\mathbb{P_{S}}(\mathbf{w}^{[d=0]}) \right \|_{2}^{2} \nonumber\\
&\leq \left \| \mathbf{w}^{[d]} -\mathbb{P_{S}}(\mathbf{w}^{[d=0]}) \right \|_{2}^{2}
-\frac{\sum_{i\in \mathcal{N_{P}}}\left \| \mathbf{w}^{[d]}-\mathbb{P_{S}}_{^{i}}(\mathbf{w}^{[d]})  \right \|_{2}^{2}}{N_{P}}
\end{align}

Therefore, we have: 
\begin{align}
&\left \| \mathbf{w}^{[d]}-\mathbb{P_{S}}(\mathbf{w}^{[d=0]}) \right \|_{2}\nonumber\\
&
\leq \left ( 1-\frac{1}{2\alpha } \right )^{d/2}\left \| \mathbf{w}^{[d=0]}-\mathbb{P_{S}}(\mathbf{w}^{[d=0]}) \right \|_{2} \label{innerspeed}   
\end{align}

Let $\eta=\left ( 1-\frac{1}{2\alpha } \right )$. Let the right hand side of (\ref{innerspeed}) $\leq\varepsilon$ and we will have the $\overline{N}$ in Proposition \ref{proposition:inner}.  
\end{proof}

Proposition \ref{proposition:inner} shows the convergence rate analysis of the inner algorithm. Then, we will consider the convergence of outer iteration. 

\begin{proposition}
\label{proposition:vas}
Use the algorithm (\ref{GD}) and (\ref{projection}), (\ref{distributed projection}) and (\ref{average projection}) to solve \eqref{problem1},   let $\mathbf{X}^{*}$ be the optimal solution. Assume $ \left \| \mathbf{X}^{[k=0]}-\mathbf{X}^{*} \right \|_{2}=R_0$.
Let Assumption \ref{assumption:si_nemp} and Assumption \ref{assumption:f_smooth} hold. Let inner iteration  (\ref{distributed projection}) and (\ref{average projection}) perform for $N_{inner}^{[k]}\geq 1$ times. Then we have:
\begin{align}
    &\left \| \mathbf{X}^{[k+1]}-\mathbb{P_{S}}(\mathbf{\widetilde{X}}^{[k]}) \right \|_{2} \nonumber\\
    &\leq  \eta^{N_{inner}^{[k]}/2}(R_0+(k+1)\frac{1}{L}\left \|\triangledown f(\mathbf{X}^{*})\right \|_{2} )
\end{align}
\end{proposition}

\begin{proof}
Since $\mathbf{X}^{*}$ is the optimal solution, we have:
$$\mathbf{X}^{*}=\mathbb{P_{S}}(\mathbf{X}^{*}-\frac{1}{L}\triangledown f(\mathbf{X}^{*}))$$

Hence,
\begin{align}
&\left \| \mathbf{\widetilde{X}}^{[k]}-\mathbf{X}^{*} \right \|_{2}\nonumber\\
&=\left \| \mathbf{\widetilde{X}}^{[k]}-(\mathbf{X}^{*}-\frac{1}{L}\triangledown f(\mathbf{X}^{*})) -\frac{1}{L}\triangledown f(\mathbf{X}^{*})\right \|_{2}\nonumber\\
&\leq \left \| \mathbf{X}^{[k]}-\frac{1}{L}\triangledown f(\mathbf{X}^{[k]})-(\mathbf{X}^{*}-\frac{1}{L}\triangledown f(\mathbf{X}^{*}))\right \|_{2}  +\frac{1}{L}\left \|\triangledown f(\mathbf{X}^{*})\right \|_{2}  
\end{align}

Since:
\begin{align}
&\left \| \mathbf{X}^{[k]}-\frac{1}{L^{[k]}}\triangledown f(\mathbf{X}^{[k]})-(\mathbf{X}^{*}-\frac{1}{L^{[k]}}\triangledown f(\mathbf{X}^{*})) \right \|_{2}^{2}\nonumber\\
&=\left \| \mathbf{X}^{[k]}-\mathbf{X}^{*} \right \|_{2}^{2}+\frac{1}{(L^{[k]})^{2}}\left \| \triangledown f(\mathbf{X}^{[k]})-\triangledown f(\mathbf{X}^{*})) \right \|_{2}^{2}\nonumber\\
&-\frac{2}{L^{[k]}}\left \langle  \triangledown f(\mathbf{X}^{[k]})-\triangledown f(\mathbf{X}^{*})),\mathbf{X}^{[k]}-\mathbf{X}^{*}\right \rangle\nonumber\\
&\leq \left \| \mathbf{X}^{[k]}-\mathbf{X}^{*} \right \|_{2}^{2}-\frac{1}{(L^{[k]})^{2}}\left \| \triangledown f(\mathbf{X}^{[k]})-\triangledown f(\mathbf{X}^{*})) \right \|_{2}^{2} 
\end{align}

Therefore,
\begin{align}
&\left \| \mathbf{X}^{[k]}-\frac{1}{L^{[k]}}\triangledown f(\mathbf{X}^{[k]})-(\mathbf{X}^{*}-\frac{1}{L^{[k]}}\triangledown f(\mathbf{X}^{*})) \right \|_{2}\nonumber\\  
&\leq \left \| \mathbf{X}^{[k]}-\mathbf{X}^{*} \right \|_{2}
\end{align}
That is:
\begin{align}
\left \| \mathbf{\widetilde{X}}^{[k]}-\mathbf{X}^{*} \right \|_{2}   
\leq \left \| \mathbf{X}^{[k]}-\mathbf{X}^{*} \right \|_{2}+\frac{1}{L}\left \|\triangledown f(\mathbf{X}^{*})\right \|_{2}\label{tilk}
\end{align}

And according Lemma \ref{lemma:inner}(b), we have:
\begin{align}
    \left \| \mathbf{w}^{[d_1]}-\mathbf{Z} \right \|_{2}\geq \left \| \mathbf{w}^{[d_2]}-\mathbf{Z} \right \|_{2}, \forall d_2\geq d_1
\end{align}

Since $\mathbf{w}^{[d=0]}=\mathbf{\widetilde{X}}^{[k]}$ and $\mathbf{w}^{[N_{inner}^{[k]}+1]}=\mathbf{X}^{[k+1]}$.

Therefore,
\begin{align}
\left \| \mathbf{\widetilde{X}}^{[k]}-\mathbf{X}^{*} \right \|_{2}   
\geq \left \| \mathbf{X}^{[k+1]}-\mathbf{X}^{*} \right \|_{2}\label{tilk+1}
\end{align}

Combine (\ref{tilk+1}) and (\ref{tilk}), we have:
\begin{align}
&\left \| \mathbf{X}^{[k+1]}-\mathbf{X}^{*} \right \|_{2}\nonumber\\
&\leq \left \| \mathbf{\widetilde{X}}^{[k]}-\mathbf{X}^{*} \right \|_{2}\nonumber\\
&\leq R_0+(k+1)\frac{1}{L}\left \|\triangledown f(\mathbf{X}^{*})\right \|_{2}     
\end{align}
Then, according to (\ref{innerspeed}), we have:
\begin{align}
 &\left \| \mathbf{X}^{[k+1]}-\mathbb{P_{S}}(\mathbf{\widetilde{X}}^{[k]}) \right \|_{2}\leq  \eta^{N_{inner}^{[k]}/2}\left \| \mathbf{\widetilde{X}}^{[k]}-\mathbb{P_{S}}(\mathbf{\widetilde{X}}^{[k]}) \right \|_{2} \label{varepsilonk}  
\end{align}

Then, let's find the bound of $\left \| \mathbf{\widetilde{X}}^{[k]}-\mathbb{P_{S}}(\mathbf{\widetilde{X}}^{[k]}) \right \|_{2}$:
\begin{align}
&\left \| \mathbf{\widetilde{X}}^{[k]}-\mathbb{P_{S}}(\mathbf{\widetilde{X}}^{[k]}) \right \|_{2}\nonumber\\
&\leq \left \| \mathbf{\widetilde{X}}^{[k]}-\mathbf{X}^{[*]} \right \|_{2}\nonumber\\
&
\leq R_0+(k+1)\frac{1}{L}\left \|\triangledown f(\mathbf{X}^{*})\right \|_{2}     
\end{align}

Combining this with (\ref{varepsilonk}) wraps up the proof.
\end{proof}

Further, we introduce Lemma \ref{lemma:outer_con} according to \cite{8290985}.
\begin{lemma}
\label{lemma:outer_con}
Use the algorithm (\ref{GD}) and (\ref{projection}) to solve \eqref{problem1}, $\mathbb{S}$ is a nonempty closed convex set. Let $\mathbf{X}^{*}$ be the optimal solution. Let Assumption \ref{assumption:f_smooth} holds. Then, $\forall k\geq 0$, we have:

(a)
\begin{align}
&\frac{1}{L^{2}}\left \| \triangledown f(\mathbf{X}^{[k]})-\triangledown f(\mathbf{X}^{*})) \right \|_{2}^{2}\nonumber\\
&\leq  \left \| \mathbf{X}^{[k]}-\mathbf{X}^{*} \right \|_{2}^{2}-\left \| \mathbf{X}^{[k+1]}-\mathbf{X}^{*} \right \|_{2}^{2}\nonumber\\
& +2\left \| \mathbf{X}^{[k+1]}-\mathbb{P_{S}}(\mathbf{\widetilde{X}}^{[k]}) \right \|_{2}
\left \| \mathbf{X}^{[k+1]}-\mathbf{X}^{*} \right \|_{2}\label{deltaf}
\end{align}

(b)
\begin{align}
&f(\frac{1}{k}\sum_{\kappa =0}^{k-1}\mathbf{X}^{[\kappa +1]})-f(\mathbf{X}^{*})\nonumber\\
& \leq\frac{L}{2k}\sum_{\kappa =0}^{k-1}(\left \| \mathbf{X}^{[\kappa ]}-\mathbf{X}^{*} \right \|_{2}^{2}-\left \| \mathbf{X}^{[\kappa +1]}-\mathbf{X}^{*} \right \|_{2}^{2}\nonumber\\
&
+2\left \| \mathbf{X}^{[\kappa +1]}-\mathbf{X}^{*} \right \|_{2}\left \| \mathbf{X}^{[k+1]}-\mathbb{P_{S}}(\mathbf{\widetilde{X}}^{[k]}) \right \|_{2})\nonumber\\
& +\frac{1}{k}\sum_{\kappa =0}^{k-1}\left \| \triangledown f(\mathbf{X}^{*}) \right \|_{2}\left \| \mathbf{X}^{[k+1]}-\mathbb{P_{S}}(\mathbf{\widetilde{X}}^{[k]}) \right \|_{2}\nonumber\\
& +\frac{1}{k}\sum_{\kappa =0}^{k-1}\left \| \mathbf{X}^{[k+1]}-\mathbb{P_{S}}(\mathbf{\widetilde{X}}^{[k]}) \right \|_{2}\left \| \triangledown f(\mathbf{X}^{[\kappa]})-\triangledown f(\mathbf{X}^{*})) \right \|_{2} \label{eq:con_rate_outer}
\end{align}
\end{lemma}

Then, we could give the convergence condition of the outer iteration. 
\begin{proposition}
\label{proposition:outer_con}
Use the algorithm (\ref{GD}) and (\ref{projection}), (\ref{distributed projection}) and (\ref{average projection}) to solve \eqref{problem1},   let $\mathbf{X}^{*}$ be the optimal solution. Let Assumption \ref{assumption:si_nemp} and Assumption \ref{assumption:f_smooth} hold. When $\underset{k\rightarrow \infty }{\lim} \frac{\log k}{N_{inner}^{[k]}}=0$, the algorithm converges to the optimal point.
\end{proposition}

\begin{proof}
By summing over the entire history of (\ref{deltaf}), we have:
\begin{align}
&(\frac{1}{k}\sum_{\kappa=0}^{k-1 }\left \| \triangledown f(\mathbf{X}^{[\kappa]})-\triangledown f(\mathbf{X}^{*})) \right \|_{2})^{2}\nonumber\\
&\leq\frac{1}{k}\sum_{\kappa=0}^{k-1 }\left \| \triangledown f(\mathbf{X}^{[\kappa]})-\triangledown f(\mathbf{X}^{*})) \right \|_{2}^{2}\nonumber\\
&\leq  L^{2}\frac{R_0^{2}-\left \| \mathbf{X}^{[k]}-\mathbf{X}^{*} \right \|_{2}^{2}}{k}\nonumber\\
 &+\frac{2 L^{2}}{k}\sum_{\kappa=0}^{k-1 }\left \| \mathbf{X}^{[\kappa+1]}-\mathbb{P_{S}}(\mathbf{\widetilde{X}}^{[\kappa]}) \right \|_{2}
\left \| \mathbf{X}^{[\kappa +1]}-\mathbf{X}^{*} \right \|_{2}
\end{align}

Let's look at the last term of the right hand side. Since:

\begin{align}
&\frac{2L^2}{k}\sum_{\kappa=0}^{k-1 }\left \| \mathbf{X}^{[\kappa+1]}-\mathbb{P_{S}}(\mathbf{\widetilde{X}}^{[\kappa]}) \right \|_{2}\left \| \mathbf{X}^{[\kappa+1]}-\mathbf{X}^{*} \right \|_{2}
\nonumber\\
&\leq  \frac{2L^2}{k}
\sum_{\kappa=0}^{k-1 }\eta^{N_{inner}^{[\kappa]}/2}(R_0+(\kappa+1)\frac{1}{L}\left \|\triangledown f(\mathbf{X}^{*})\right \|_{2})^2
\end{align}

The sequence $\left \{ \frac{1}{k}\sum_{\kappa=0}^{k-1 }\left \| \triangledown f(\mathbf{X}^{[\kappa]})-\triangledown f(\mathbf{X}^{*})) \right \|_{2} \right \}$ converges to zero when 
\begin{align}
&\underset{k\rightarrow \infty }{\lim}\frac{1}{k}\sum_{\kappa=0}^{k-1 }\eta^{N_{inner}^{[\kappa]}/2}(R_0+(\kappa+1)\frac{1}{L}\left \|\triangledown f(\mathbf{X}^{*})\right \|_{2})^2=0
\end{align}

According to Stolz-Cesaro theorem, we need:
$\underset{k\rightarrow \infty }{\lim} \frac{\log k}{N_{inner}^{[k]}}=0$.
\end{proof}

Also, we could have the proposition for the convergence rate of outer iteration as below.
\begin{proposition}
\label{proposition:outer_rate}
Use the algorithm (\ref{GD}) and (\ref{projection}), (\ref{distributed projection}) and (\ref{average projection}) to solve \eqref{problem1},   let $\mathbf{X}^{*}$ be the optimal solution. Let Assumption \ref{assumption:si_nemp} and Assumption \ref{assumption:f_smooth} hold. We assume that $N_{inner}^{[k]}=N_0, k\geq1$ and $\left \| \triangledown f(\mathbf{X}^{*}) \right \|_{2}\leq LR_0$. Then, given $\forall \epsilon >0$,
after at most $N_{max}=\overline{k}\bar{N}$ steps, that is, after $\overline{k}$ outer iterations where each performs $\bar{N}$ inner iterations, we could obtain a $\frac{1}{k}\sum_{\kappa =0}^{k-1}\mathbf{X}^{[\kappa +1]}$ satisfying: $f(\frac{1}{k}\sum_{\kappa =0}^{k-1}\mathbf{X}^{[\kappa +1]})-f(\mathbf{X}^{*})<\epsilon$, where: \begin{align}
& \overline{k}=\max\left[ \frac{2LR_0}{\varepsilon},\frac{\sqrt{2}L^2R_0^2}{\sqrt{\varepsilon}}, 1\right] \nonumber\\
&\bar{N}=\max\left[ \frac{2\log\left( \frac{8\sqrt{2} LR_0^3}{\varepsilon\sqrt{\varepsilon}}(L^5R_0^3+L^2+1)\right)}{\log\eta}, 1\right.\nonumber\\
&\frac{2\log\left( \frac{8\sqrt{2}L^6R_0^6}{\varepsilon\sqrt{\varepsilon}}+\frac{24L^4R_0^4}{\varepsilon}+\frac{8\sqrt{2}L(L^2+1)R_0^3}{\varepsilon\sqrt{\varepsilon}}+\frac{4LR_0^2}{\varepsilon}+\frac{(8L^2+4)R_0}{\varepsilon L}\right)}{\log\eta},\nonumber\\
&\frac{\log\left( \frac{8L^2(2L^2+1)R_0^4}{\varepsilon^3}+\frac{16L(L^2+1)R_0^3}{\varepsilon^2}+\frac{8(L^2+1)R_0^2}{\varepsilon}\right)}{\log\eta},\nonumber\\
&\left.
\frac{\log\left( \frac{4L^4(2L^2+1)R_0^6}{\varepsilon^2}+\frac{8\sqrt{2}L^2(L^2+1)R_0^4}{\varepsilon\sqrt{\varepsilon}}+\frac{8(L^2+1)R_0^2}{\varepsilon}\right)}{\log\eta}\right]
\end{align}
\end{proposition}

\begin{proof}
According to \eqref{eq:con_rate_outer} and $\left \| \triangledown f(\mathbf{X}^{*}) \right \|_{2}\leq L\left \| \mathbf{X}^{[k=0 ]}-\mathbf{X}^{*} \right \|_{2}$, we get:
\begin{align}
&f(\frac{1}{k}\sum_{\kappa =0}^{k-1}\mathbf{X}^{[\kappa +1]})-f(\mathbf{X}^{*})\nonumber\\
&\leq\frac{LR_0}{2k}+\frac{L^4R_0^4}{2k^2}\nonumber\\
&+\eta^{N_0/2}\left( R_0^4(k+3)2L^4+R_0^2L+R_0\left( L(2+k)+\frac{k+1}{L} \right) \right)\nonumber\\
&+\eta^{N_0}R_0^2\left( k^2(L^2+\frac{1}{2})+2(k+1)(L^2+1) \right)\label{eq:con_rate_outer}
\end{align}

Let each of the four terms at the right hand of \eqref{eq:con_rate_outer} no larger than $\epsilon/4$. Then, we have the upper bound of $\overline{k}$ with $\bar{N}$ as in Proposition \ref{proposition:outer_rate}.
Multiply $\overline{k}$ with $\bar{N}$, we could get Proposition \ref{proposition:outer_rate}.
\end{proof}
Till here, we have presented the convergence analysis of the proposed method; the \textbf{convergence condition} in Proposition \ref{proposition:outer_con} and the \textbf{convergence rate} in Proposition \ref{proposition:outer_rate}.
\subsection{Parameter design analysis}
According to section 4.2, we will present a parameter design analysis of the proposed method. 
\subsubsection{Learning rate}
The learning rate $1/L$ should satisfy Assumption \ref{assumption:f_smooth} to ensure the convergence of the proposed method. Therefore, $L$ should be larger than any quadratic term coefficient of the objective function in problem \eqref{problem1}. That is:
\begin{align}
    L\leq\max\left[ \lambda_{DE1}^{i,p},\beta_{PL2}^{i,m},\forall i \in \mathcal{N_{P}},\forall m \in \mathcal{N}_{PL}^{i},\forall p \in \mathcal{N}_{DE}^{i}\right] \label{eq:learning_rate}
\end{align}

$\beta_{PL2}^{i,m}$ indicates the sensitivity towards the total power consumption. It is usually bounded in a range where prosumers could choose one value for themselves.
$\lambda_{DE1}^{i,p}$ is the oil cost coefficient of diesel engines. The oil prices determine the upper bound in history. Therefore, as long as we set the learning rate according to \eqref{eq:learning_rate}, the convergence is guaranteed. Then, as it works in traditional gradient descent methods, a large learning rate speeds up the algorithm whereas a smaller rate improves the accuracy.
\subsubsection{Inner iteration times}

According to Proposition \ref{proposition:vas}, large $N_{inner}^{[k]}$ results in a more accurate projection. With the $N_{inner}^{[k]}$ large enough, the algorithm tends to become the exact projection gradient descent. We could fix $N_{inner}^{[k]}=N_0$ as a large number in the simplest way. Then, the convergence is guaranteed according to Proposition \ref{proposition:outer_rate}. However, large $N_{inner}^{[k]}$ will also slow down the algorithm. 

To sum up, there is a trade-off between accuracy and speed when designing the parameters of the proposed method. However, convergence could be guaranteed at the expense of speed.

\subsection{Decentralized realization}
This subsection will derive a decentralized representation of the proposed inner-outer iteration algorithm. We will prove that each prosumer only needs to communicate with its neighboring agents (prosumers).
\subsubsection{Definition of neighbor}
To facilitate analysis, let $r$ be the index number of the $r$th element in $\mathbf{X}$ and $\mathbf{X}^i$ be the vector of all the stacked variables of Prosumer $i$. That is: $\mathbf{X}=[x_1...x_r...]^T=[{\mathbf{X}^1}^T...{\mathbf{X}^i}^T...]^T$. Then, we divide the variables into two types: \textbf{\textit{uncoupled variables}}: $P_{PL}^{i,m}(t)$, $P_{D}^{i,q}(t)$, $ P_{C}^{i,q}(t)$, $P_{DE}^{i,p}$, $P_{gs}^{i}(t)$, $P_{gb}^{i}(t)$, and \textbf{\textit{coupled variables}}: $P_{ps}^{i,j}(t)$, $P_{pb}^{i,j}(t)$, $P_{o}^{i}(t)$. Let $\mathbf{X}^{i}_{u}$ be the vector of the uncoupled stacked variables of Prosumer $i$, 
and $\mathbf{X}^{i}_{c}$ denotes the vector of all the stacked coupled variables of Prosumer $i$, i. e., $\mathbf{X}^i={[{\mathbf{X}^{i}_{u}}^T,{\mathbf{X}^{i}_{c}}^T]}^{T}$. 

Any element $x_{r},x_{r}\in \mathbf{X}^{i}_{c}$ is coupled with some other elements $x_{\bar{r}},x_{\bar{r}}\in \mathbf{X}^{j}_{c}, j\neq i$ by \eqref{CPpsb} or \eqref{Cline}. Then, Prosumer $i$ and Prosumer $j$ are \textbf{\textit{neighbors}} related to $x_{r}$ or $x_{\bar{r}}$. 
Let $\mathcal{N}_{c}^{r}$ denote the set of neighbors related to $x_{r}$.  For instance, consider $x_{r}=P_{pb}^{1,2}(0)$, this variable only refers to 
 prosumer $2$ who has variables $P_{ps}^{2,1}(0)$ coupled with $x_{r}$. Hence, $\mathcal{N}_{c}^{r}=\left \{ 1, 2 \right \}$. We will show later in next section that Prosumer $i$ needs to share the updated value of $x_{r}$ with any prosumer $j\in \mathcal{N}_{c}^{r}$.

Then, the Proposition \ref{network} can be stated as below. Proposition \ref{network} shows that \textit{one's neighbors related to some coupled variable} and \textit{his neighbors' neighbors related to this coupled variable} are the same. 
\begin{proposition}
\label{network}
For the energy model defined in (\ref{problem1}),  $\forall x_{r},x_{r}\in \mathbf{X}^{i}_{c}$, assume  $x_{\bar{r}},x_{\bar{r}}\in \mathbf{X}^{j}_{c}, j\neq i$ is coupled with $x_{r}$ through \eqref{CPpsb} or \eqref{Cline}.
Then $\mathcal{N}_{c}^{r}=\mathcal{N}_{c}^{\bar{r}}$. 
\end{proposition}
\begin{proof}
$\forall x_{r}, x_{r}\in \mathbf{X}^{i}_{c}$ is either $P_{ps}^{i,j}(t)$, $P_{pb}^{i,j}(t)$ or $P_{o}^{i}(t)$. 

\noindent 1) When $x_{r}$ is $P_{ps}^{i,j}(t)$ or $P_{pb}^{i,j}(t)$, $x_{\bar{r}}$ is either $P_{ps}^{j,i}(t)$ or $P_{pb}^{j,i}(t)$. Prosumer $i$ and $j$ are the only two coupled prosumers of these elements, and $\mathcal{N}_{c}^{r}=\mathcal{N}_{c}^{\bar{r}}=\left \{ i,j \right \}$.  

\noindent 2) When $x_{r}$ is $P_{o}^{i}(t)$, $x_{\bar{r}}$ is $P_{o}^{j}(t)$, where $j$ satisfies $\exists r, r\in \mathcal{L}, Y_{l}^{i}= Y_{l}^{j}=1, $. Consider the physical meaning of $\eqref{Cline}$. Assume there exists another line $\bar{l}$ ($\bar{l}\neq l$) and another prosumer $g$ ($g\neq j,i$),  and $Y_{\bar{l}}^{j}= Y_{\bar{l}}^{g}=1$.  Then, $Y_{\bar{l}}^{i}=1$. Otherwise, Prosumer $j$ is connected to a child node with two parent nodes. Therefore, $g\in \mathcal{N}_{c}^{r},g\in \mathcal{N}_{c}^{\bar{r}}$. That is, $\mathcal{N}_{c}^{r}=\mathcal{N}_{c}^{\bar{r}}$.
\end{proof}
 \subsubsection{Updating variables}In the energy sharing model, Prosumer $i$ only has access to $\mathbb{S}^{i}$, $\mathbf{X}^{i}$ and $\mathcal{N}_{c}^{r}$, $\forall x_{r}\in \mathbf{X}^{i}_{c}$. In this paper, let Prosumer $i$ also have access to those elements $x_{\bar{r}},x_{\bar{r}}\in \mathbf{X}^{j}_{c}, j\neq i, j\in \mathcal{N}_{c}^{r}$. According to Proposition \ref{network}, we will show that the inner-outer iteration method can be done in a decentralized way by communicating with neighbors. The following procedure is the foundation our decentralized decision making approach.

\textbf{Updating \eqref{GD}}: By observing $f(\mathbf{X})$ in \eqref{problem1}, we find that there is no coupled relationship in $\triangledown f(\mathbf{X})$. Put differently, $\forall x_{r}\in \mathbf{X}^{i}$, $\frac{\partial f^i}{\partial x_r}$ is only determined by $ x_r$. Therefore, for every outer iteration, Prosumer $i$ just calculates the following:

\begin{equation}
  \tilde{x}_r^{[k+1]}={x}_r^{[k]}-\frac{1}{L}\frac{\partial f^i}{\partial {x}_r^{[k]}}\label{GDdis}  
\end{equation}

 Then, each prosumer will hold his own part of $\mathbf{\tilde{X}}^{[k]}$. After \eqref{GDdis}, let each prosumer send the updated value of coupled variables to the corresponding neighbors. That is, for every $r$($\forall x_{r}\in \mathbf{X}^{i}_c$), Prosumer $i$ should send $\tilde{x}_r^{[k+1]}$ to Prosumer $j$($j\in \mathcal{N}_{c}^{r}$). 

\textbf{Updating \eqref{distributed projection}}:
 Equation (\ref{distributed projection}) projects onto $\mathbb{S}^{i}$. $\mathbb{S}^{i}$ is merely related to $\mathbf{X}^{i}$ and some coupled variables $x_{\bar{r}}$. Other elements will remain the same during projection. Since they won't affect the projection, Prosumer $i$ doesn't need this information. Given  the updated value of coupled variables  sent to other prosumers after \eqref{GDdis}, Prosumer $i$ is able to project onto $\mathbb{S}^{i}$. 

After \eqref{distributed projection}, Prosumer $i$ has the updated value of the following elements in $\mathbf{X}^{[d]}_{\mathbb{S}^{i}}$:
(a) Uncoupled variables of Prosumer $i$, i.e., $x_{r}, x_{r}\in \mathbf{X}^{i}_{u}$.
(b) Coupled variables of Prosumer $i$, i.e., $x_{r}, x_{r}\in \mathbf{X}^{i}_{c}$.
(c) A subset of coupled variables associated with Prosumer $j$: $x_{\bar{r}}, x_{\bar{r}}\in \mathbf{X}^{j}_{c}$, $j \in \mathcal{N}_{c}^{r}/i$, $\forall x_{r}\in \mathbf{X}^{i}_{c}$. Let Prosumer $i$ send the updated value of (b) and (c) to the corresponding neighbor after \eqref{distributed projection}. According to Proposition \ref{network}, Prosumer $i$ actually sends the updated value to its neighbors and its neighbors' neighbors.

\textbf{Updating \eqref{average projection}}:
Through (\ref{average projection}) Prosumer $i$ calculates the average. We use $w^{[d]}_{r}$ to denote $r$th element in $\mathbf{w}^{[d]}$ and $x^{[d]}_{r,\mathbb{S}^{j}}$ in $\mathbf{X}^{[d]}_{\mathbb{S}^{j}}$. According to \eqref{average projection}, we have $w^{[d+1]}_{r}=
x^{[d]}_{r,\mathbb{S}^{i}}+\sum_{j \in \mathcal{N_{P}}/i}x^{[d]}_{r,\mathbb{S}^{j}}/N_{P}$. There are four possibilities for $r$:

\noindent(i) ${x}_{r}\in \mathbf{X}^{i}_{u}$. This  uncoupled variable remains the same when other prosumers project onto $\mathbb{S}^{j}$, i.e., $x^{[d]}_{r,\mathbb{S}^{j}}=w^{[d]}_{r}$.
Therefore:

\begin{align}
w^{[d+1]}_{r}=( {(N_{P}-1)w^{[d]}_{r}}+ x^{[d]}_{r,\mathbb{S}^{i}})/N_{P}
\end{align}

(ii) ${x}_{r}\in \mathbf{X}^{i}_{c}$. According to proposition \ref{network}, Prosumer $i$ could receive $x^{[d]}_{r,\mathbb{S}^{j}}$ from prosumer $j,\forall j \in \mathcal{N}_{c}^{r}/i$. ${x}_{r}$ remains the same when the other prosumers $g\notin \mathcal{N}_{c}^{r}$ project onto $\mathbb{S}^{g}$.
Hence,

\begin{align}
    w^{[d+1]}_{r}=((N_{P}-N_{c}^{r})w^{[d]}_{r}+ x^{[d]}_{r,\mathbb{S}^{i}} +\sum_{j \in \mathcal{N}_{c}^{r}/i}x^{[d]}_{r,\mathbb{S}^{j}}  )/N_{P}
\end{align}

(iii) ${x}_{r}$ is in $\mathbf{X}^{j}_{c}, j\neq i$ and is coupled with Prosumer $i$'s variables by \eqref{CPpsb} or \eqref{Cline}. Then, Prosumer $i$ could receive $x^{[d]}_{r,\mathbb{S}^{g}}$ from prosumer $g,\forall g \in \mathcal{N}_{c}^{r}/i$ according to proposition \ref{network}. ${x}_{r}$ remains the same when the other prosumers $h, h\notin \mathcal{N}_{c}^{r}$ project onto $\mathbb{S}^{h}$.
Therefore:

\begin{align}
 w^{[d+1]}_{r}= ((N_{P}-N_{c}^{l})w^{[d]}_{r}+ x^{[d]}_{r,\mathbb{S}^{i}} +\sum_{g \in \mathcal{N}_{c}^{r}/i}x^{[d]}_{r,\mathbb{S}^{g}}   )/N_{P}   
\end{align}

(iv) ${x}_{r}$ is in $\mathbf{X}^{j}_{u}, j\neq i$, or it is in $\mathbf{X}^{j}_{c},j\neq i$ but $j$ is not a neighbor of Prosumer $i$ related to any variable. ${x}_{r}$ won't affect the projection onto $\mathbb{S}^{i}$, and Prosumer $i$ is not required to update this variable.

\begin{figure}[htbp]
\centering
\setlength{\abovecaptionskip}{0.cm}
\includegraphics[width=0.99\columnwidth]{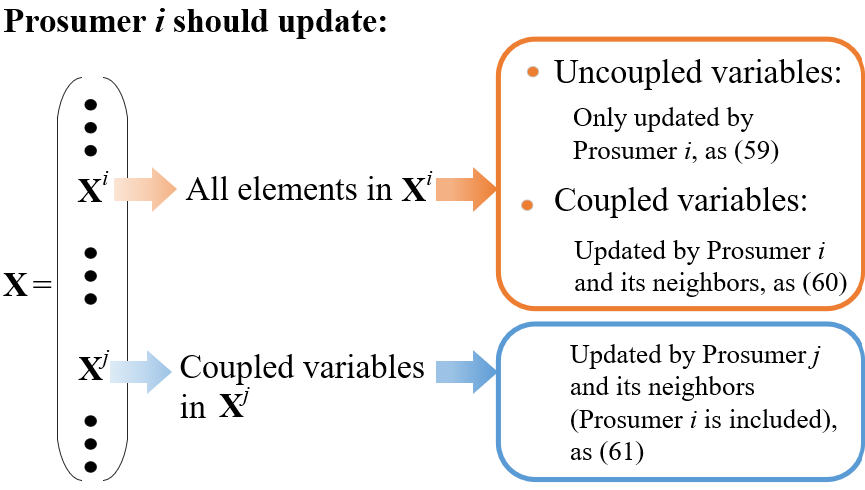}
\caption{Updating variables.}
\label{f:update}
\end{figure}

Note, by taking the outlined steps to update ${x}_{r}$, the algorithm reduces to \eqref{average projection} and can be implemented in a decentralized manner, as summarized in Figure \ref{f:update}.
\section{Study case}
\begin{figure}[htbp]
\centering
\setlength{\abovecaptionskip}{0.cm}
\includegraphics[width=0.9\columnwidth]{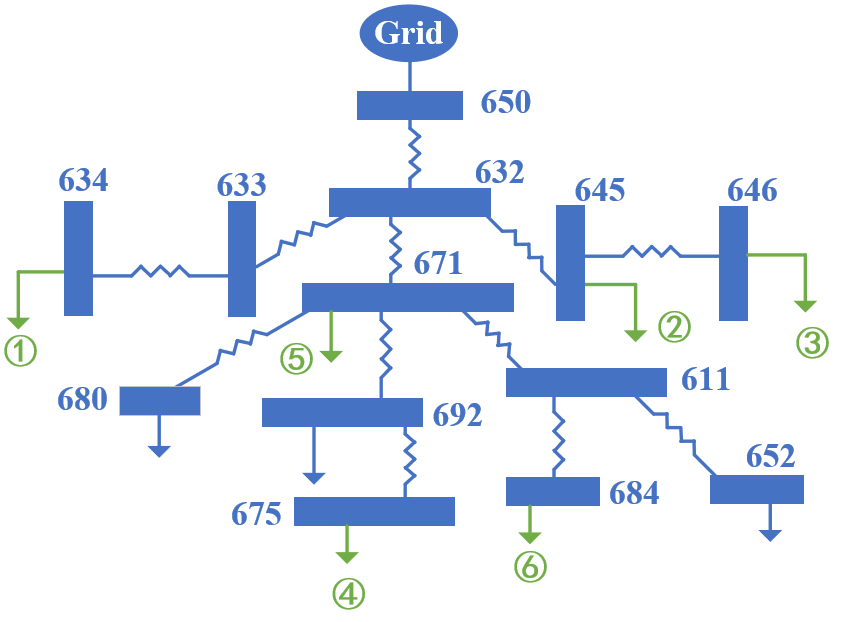}
\caption{IEEE 13bus system.}
\label{f:system}
\end{figure}

The IEEE 13 bus system (shown in Figure \ref{f:system}) is used to verify the proposed method. We leverage the proposed algorithm to determine the quantity of traded energy. Only the results for problem \eqref{problem1} is presented here. We assume each line in the system shares the same electrical distance for simplicity of \eqref{eq:electrical_d}. 

\begin{figure}[htbp]
\centering
\setlength{\abovecaptionskip}{0.cm}
\includegraphics[width=1\columnwidth]{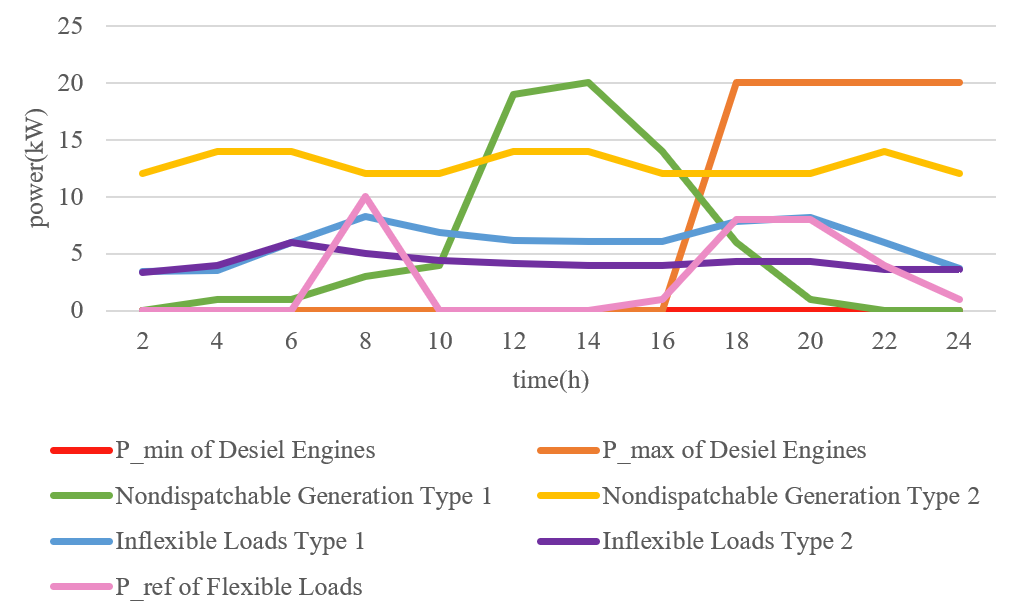}
\caption{The power profile of prosumers.}
\label{f:profile}
\end{figure}

Assume $N_p=6$ and each prosumer has one flexible load, one ESS. Only Prosumer 5 and Prosumer 6 have diesel engines. The inflexible load profile of Prosumer 1 and Prosumer 2 is as type 1 in Figture.\ref{f:profile} and they don't have nondispatchable generators. The load profile of other prosumers is as type 2 in Figture.\ref{f:profile}. The nondispatchable generation profile of Prosumer 3 and Prosumer 5 is as type 1 and type 2 for Prosumer 4 and Prosumer 6. The parameters are presented in Table \ref{table:para}. The unit of power is $kW$.

\begin{table}[htbp]
\caption{Generation and loads parameters}
\begin{center}
\label{table:para}
\begin{tabular}{|c|c|c|c|}
\hline
\textbf{Parameters}           & \textbf{Value}    & \textbf{Parameters}           & \textbf{Value}      \\ \hline
$P_{PLmin,t}$        & 0         & $R_{up,t}^{i,p}$      & -2 \\ \hline
$P_{PLmax,t}$        & 10        & $R_{down,t}^{i,p}$    & 2 \\ \hline
$W_{PLref}^{i,m}$    & 40        & $\lambda_{DE1}^{i,p}$       & 0 \\ \hline
$\beta_{PL1}^{i,m} $ & 0.01 & $\lambda_{DE2}^{i,p}$       & 0.2214 \\ \hline
$\beta_{PL2}^{i,m} $ & 0.01 & $\Delta t$            & 2          \\ \hline
$P_{Cmax}^{i,q}$     & 10 & $W_{max}^{i,q}$       & 5 \\ \hline
$P_{Dmax}^{i,q}$     & 10 & $\eta _{C}^{i,q}$     & 0.95 \\ \hline
$W_{ESSN}^{i,q}$ &     50 & $\lambda _{o}$
   & 0.01\\ \hline
$SOC_{min}^{i,q}$    & 0.15 & $\eta _{D}^{i,q}$     & 0.95 \\ \hline
$SOC_{max}^{i,q}$    & 0.85 & $W_{ESS0}^{i,q}$     & 25 \\ \hline
$W_{min}^{i,q}$      & -5 & $\lambda _{ESS}^{i,q}$ & 0.1 \\ \hline
$\lambda _{gs}$      & 0.1 &$\lambda _{gb}$ &0.23 \\ \hline
\end{tabular}
\end{center}
\end{table}

Assume $1/L=100,N_{inner}^{[k]}=100$. Then, there are $N_x=1176$ elements in $\mathbf{X}$. We use zero for variables' initial values. 
\subsection{Distributed realization}

\begin{figure}[htbp]
\centering
\setlength{\abovecaptionskip}{0.cm}
\subfloat[Error of power]{\includegraphics[width=1\columnwidth]{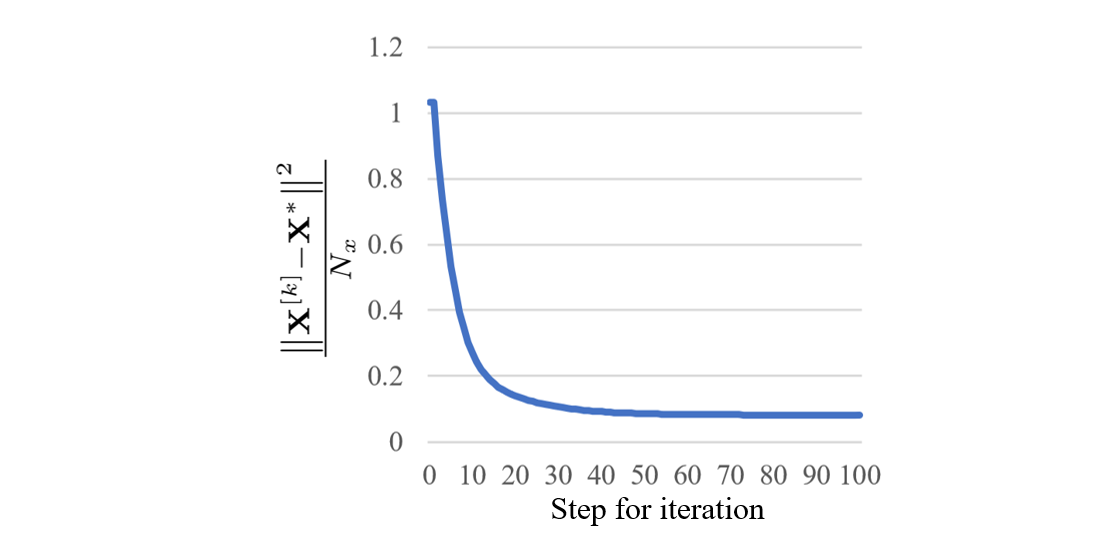}}
\newline
\subfloat[Error of the cost function]{\includegraphics[width=1\columnwidth]{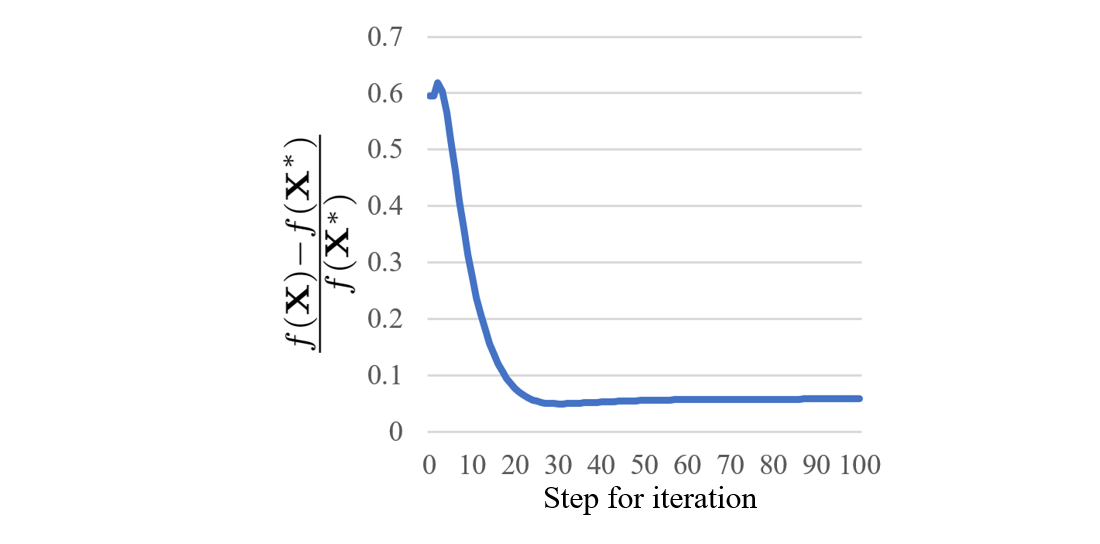}}
\caption{The error of the decentralized inexact projection method.}
\label{f:distributed}
\end{figure}

The error $\frac{\left \| \mathbf{X}^{[k]}-\mathbf{X}^{*} \right \|^{2}}{N_x}$ and the accuracy $\frac{f(\mathbf{X})-f(\mathbf{X}^{*})}{f(\mathbf{X}^{*})}$ are given as Figure  \ref{f:distributed}.
After 100 iterations, the average error of each variable is 0.08 kW, and the error of the objective function is 0.058\%.

\subsection{Energy trading}
\begin{figure}[]
\centering
\setlength{\abovecaptionskip}{0.cm}
\includegraphics[width=1\columnwidth]{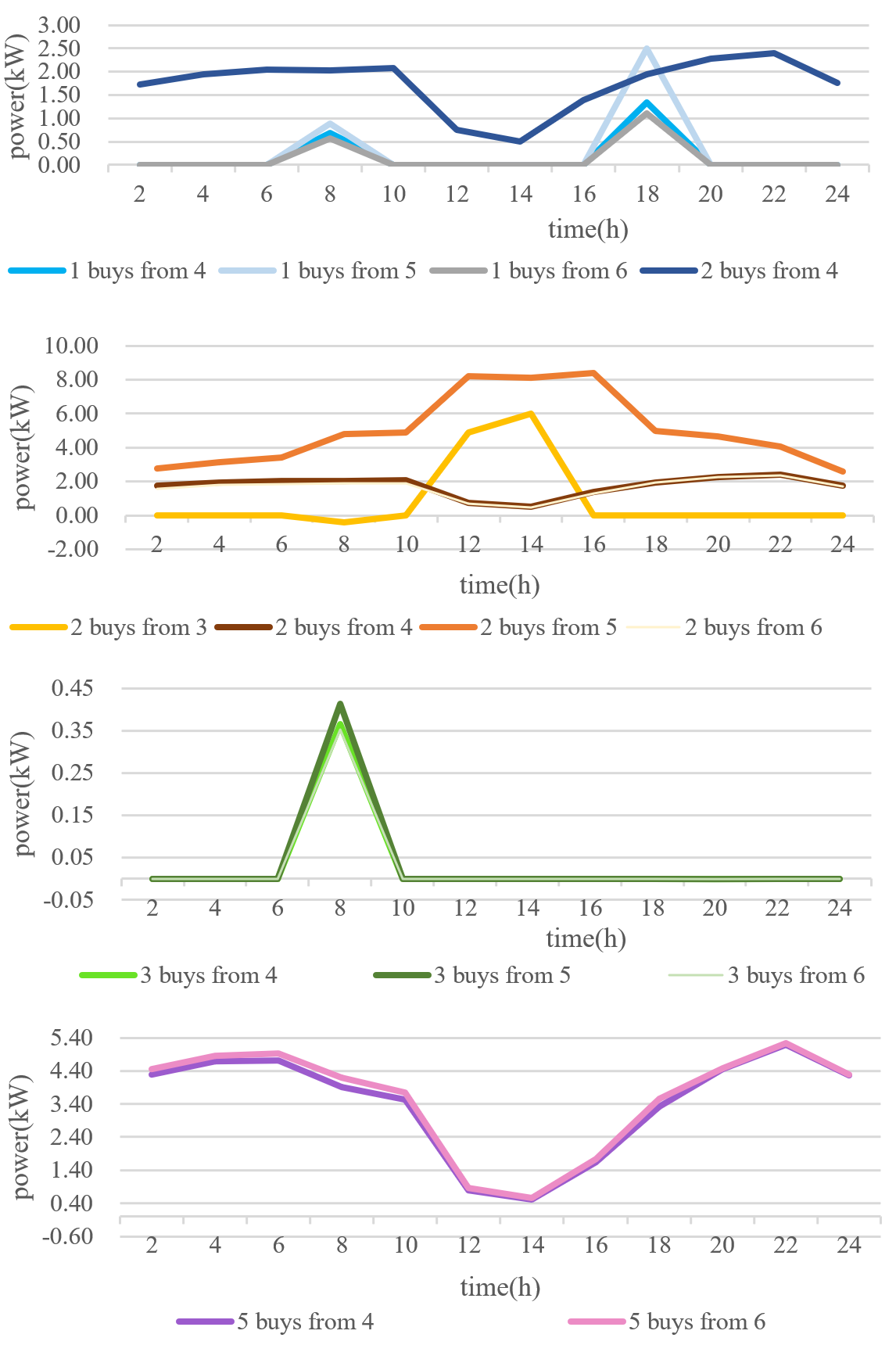}
\caption{Energy trading among prosumers.}
\label{f:energysharing}
\end{figure}

Figure \ref{f:energysharing} shows the energy trading profile among prosumers. Most energy sharing happens between 10:00 and 18:00 when Prosumer 3-6 has excess energy. Prosumer 5 serves as a buyer in the morning and at night and becomes a seller around noon. P2P energy trading allows players' flexibility as long as the total social cost is minimized. Since prosumer 5 is closer to prosumer 4, he has the highest priority for prosumer 4. Similarly, prosumer 2 prefers to trade energy with prosumer 5 instead of prosumer 6.

\subsection{Demand-supply relationship}
\begin{figure}[]
\centering
\setlength{\abovecaptionskip}{0.cm}
\includegraphics[width=1\columnwidth]{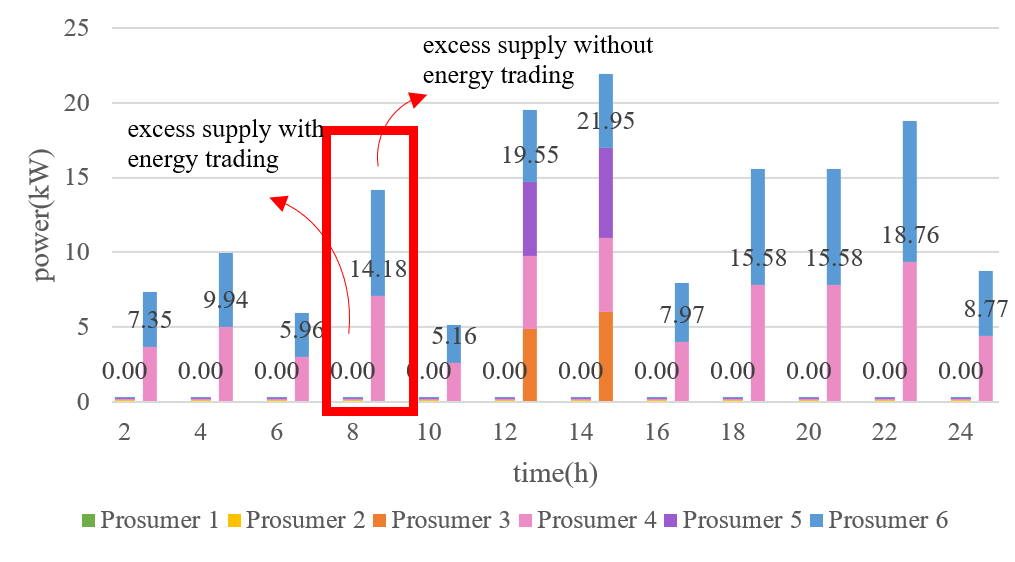}
\caption{Excess supply with\&without energy trading.}
\label{f:supply}
\end{figure}

\begin{figure}[]
\centering
\setlength{\abovecaptionskip}{0.cm}
\includegraphics[width=1\columnwidth]{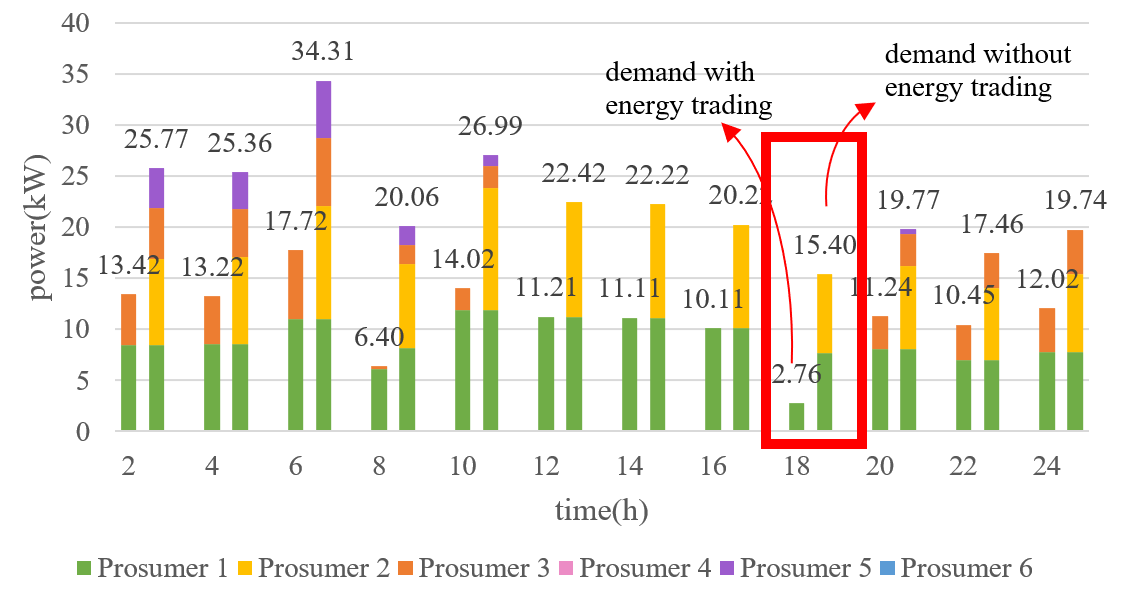}
\caption{Demand with\&without energy trading.}
\label{f:demand}
\end{figure}

Figure \ref{f:supply} and \ref{f:demand} shows the demand-supply relationship with and without energy trading. As it can be seen, the P2P community has less energy congestion in both supply and demand with energy trading. Excess supply can be stored or consumed in the community so that the community's demand also decreases. Energy trading could alleviate the demand-supply congestion.

\section{Conclusion}
This paper presents a fully decentralized inexact projection method to solve Peer-to-Peer energy trading problems.  Each prosumer only needs to share the updated coupled variable with neighboring prosumers. The performance of our approach does not rely on the tuning of the hyperparameters, which addresses a significant drawback of decentralized methods. The simulation
results based on IEEE 13 bus system show the convergence of the algorithms and the effectiveness of the proposed solution to solve the P2P energy
sharing problem. The results also show that P2P energy trading contributes to supply-demand equity and alleviates congestion in communities.


\bibliographystyle{ACM-Reference-Format}
\bibliography{sample-base}


\end{document}